\documentclass[runningheads]{llncs}
\usepackage{graphicx}
\usepackage{tcolorbox}
\usepackage{multirow}
\usepackage{amsmath}
\usepackage[ruled,linesnumbered]{algorithm2e}

\begin{document}

\title{On Unified Prompt Tuning for Request Quality Assurance in Public Code Review}
\titlerunning{Request Quality Assurance in Public Code Review}

\author{
Xinyu Chen\inst{1} \and
Lin Li\inst{1*}     \and
Rui Zhang\inst{1} \and
Peng Liang\inst{2}}
\authorrunning{Chen et al.}

\institute{School of Computer Science and Artificial Intelligence, Wuhan University of Technology, Wuhan 430070, China\\
\and
School of Computer Science, Wuhan University, Wuhan 430072, China\\
\email{\{268595, cathylilin, zhangrui\}@whut.edu.cn, liangp@whu.edu.cn}}

\maketitle             
\begin{abstract}
Public Code Review (PCR) can be implemented through a Software Question Answering (SQA) community, which facilitates high knowledge dissemination. 
Current methods mainly focus on the reviewer's perspective, including finding a capable reviewer, predicting comment quality, and recommending/generating review comments. 
Our intuition is that satisfying review necessity requests can increase their visibility, which in turn is a prerequisite for better review responses.
To this end, we propose a unified framework called UniPCR to complete developer-based request quality assurance (i.e., predicting request necessity and recommending tags subtask) under a Masked Language Model (MLM). 
Specifically, we reformulate both subtasks via
1) text prompt tuning, which converts two subtasks into MLM by constructing prompt templates using hard prompt; 2) code prefix tuning, which optimizes a small segment of generated continuous vectors as the prefix of the code representation using soft prompt.
Experimental results on the Public Code Review dataset for the time span 2011-2023 demonstrate that our UniPCR framework adapts to the two subtasks and outperforms comparable accuracy-based results with state-of-the-art methods for request quality assurance. 
These conclusions highlight the effectiveness of our unified framework from the developer's perspective in public code review.

\end{abstract}
\section{Introduction}

%% 1. Code review
Code review can be conducted either within the team or by public developers~\cite{DBLP:journals/ese/EistyC22}. 
Team code review primarily focuses on identifying defects, ensuring the need for code adjustments, and aiming to enhance the efficiency of the review process~\cite{DBLP:journals/corr/abs-2308-11148,DBLP:conf/icml/ZhangYHLYF023}.
% ~\cite{DBLP:journals/corr/abs-2308-11148,DBLP:conf/icml/ZhangYHLYF023}. 
Public Code Review (PCR) serves as a complementary approach, intended to improve knowledge sharing and expertise among relevant professionals during the review process~\cite{DBLP:conf/icsm/BaumSB17}.

Different from code review within the team, PCR is developed in the Software Question Answering (SQA) community, thus researchers are increasingly exploring ways to achieve high-quality and efficient review services~\cite{DBLP:conf/sigsoft/HongTTA22}. 
Recent studies have focused on code review within teams such as reducing review time~\cite{DBLP:conf/sigsoft/ShanSHRCPRN22} and recommending suitable reviewers~\cite{DBLP:conf/icse/ChenKBSXL22,li2023code}. Other studies aim to normalize reviewer responses by predicting the quality of review comments~\cite{DBLP:conf/msr/RahmanRK17,DBLP:conf/sigsoft/Pandya022} to obtain better review responses. The common goal of these works is to optimize the review process for better team code review service. 

% 2
In the PCR process implemented by the SQA community, request quality assurance mainly depends on developers.
As developers are required and encouraged to submit necessary PCR requests in the community, the necessity of these requests is evaluated by public via {\bf{request necessity prediction subtask}} after they are posted. %More details about PCR service processes are given in Section~\ref{process}.
Meanwhile, developers always need to choose a few technical terms (e.g., \textit{c}, \textit{pthreads}) as review tags. These tags can be used to match requests with suitable reviewers via {\bf{tag recommendation subtask}} (see the detailed PCR process in Section~\ref{process}). 

However, when the above subtasks need to be processed, prior researches usually fine-tuned~\cite{DBLP:conf/acl/LiL20} modelling framework for each subtask. This means that task-specific learning frameworks require additional costs to adapt the model to different subtasks, which is contrary to the demand for simple and fast tools in the field of software engineering~\cite{DBLP:conf/sigsoft/FuM17}. Inspired by prompt tuning~\cite{DBLP:journals/csur/LiuYFJHN23}, we aim to bridge this gap and improve request quality assurance performance by reformulating the traditional discriminative learning widely used in the request necessity prediction subtask and the tag recommendation subtask as generative learning.
To this end, we propose a unified prompt tuning that allows us to learn and train both subtasks for request quality assurance.
Specifically, we reformulate these discriminative subtasks via 1) text prompt tuning, % ~\cite{DBLP:conf/eacl/SchickS21}
which converts them into generative learning by constructing prompt templates using hard prompt (see Section~\ref{text prompt section}). The description of the subtasks prompts the model to learn the subtasks; and 2) code prefix tuning,
% ~\cite{DBLP:conf/acl/LiL20}
which optimizes a small segment of continuous vectors in the entire code feature space (see Section~\ref{code prompt section}). Using this short prefix to further optimize the representation learning of the code.

The \textbf{contributions} of this paper can be summarized as follows:
1) We propose a \textbf{Uni}fied framework for \textbf{P}ublic \textbf{C}ode \textbf{R}eview (UniPCR) to consider the different subtasks of request quality assurance in the public code review service. The UniPCR framework completes the unified training of both subtasks through the designed prompt template.
% \item{}We reconstruct request necessity prediction and tag recommendation subtasks based on Mask Language Model (MLM) via hard prompt for text with designed prompt templates, and soft prompt for code segments with task-specific prefixes.

2) Experimental results on the Public Code Review dataset for the time span 2011-2023 demonstrate that UniPCR outperforms comparable accuracy-based results with state-of-the-art methods for the request quality assurance.%These conclusions highlight the good performance of our unified framework from both the developer and reviewer perspectives. % 

\section{Background} \label{preliminary}

\subsection{Public Code Review Process}\label{process}
\begin{figure}[h]
  \centering
  \includegraphics[width=0.9\linewidth]{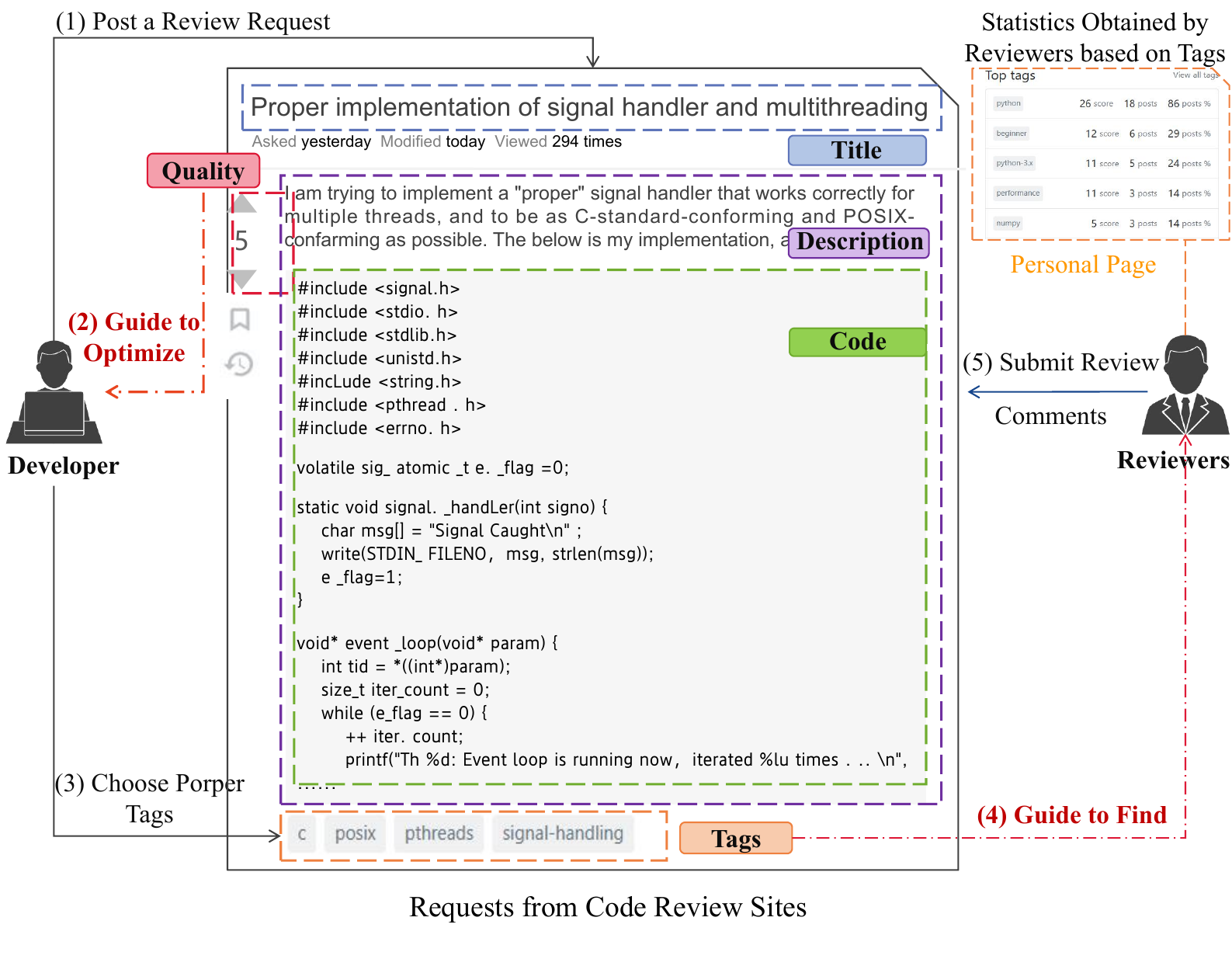}
  \caption{An Example of Public Code Review Process (with a data sample in the middle}
  \label{public example}
\end{figure}

In this section, we outline how these request quality assurance subtasks (request necessity prediction and tag recommendation subtask) serve the public code review process~\cite{DBLP:journals/ese/EistyC22}. 
The process is comprised of five core steps (as depicted in Figure~\ref{public example}), starting with the code writer submitting a request for change or optimization in Step \textcircled{1}. In Step \textcircled{2}, practitioners in the community browse and evaluate the request to assign a necessity feedback. Compliance with community norms is essential for requests to receive a response, and unnecessary requests may hinder visibility.
Subsequently, in Step \textcircled{3}, developers are asked to select several relevant technical terms to accompany their request. These tags are then used in Step \textcircled{4} to match the request with appropriate reviewers, who provide their feedback on the code in Step \textcircled{5}.  

Our task logic involves integrating the necessity prediction with Steps \textcircled{2} to provide practitioners with necessity feedback in advance. This can guide them in optimizing their request expression and improving request visibility while reducing the time spent by reviewers in Step \textcircled{5}. Additionally, the tag recommendation is associated with Steps \textcircled{4}, allowing requests to be correctly matched with appropriate reviewers based on the use of tags within the PCR community.

\vspace{-3em}

\subsection{Request Quality Assurance Task Definition} \label{taskdef}
The effectiveness of the PCR process hinges on the interaction between developers and reviewers. Unfortunately, this interaction can be time-consuming due to the need to identify issues, find appropriate reviewers, etc. While current approaches focus on reducing time costs for reviewers, they overlook the impact of developers on the review process. Our intuition is that well-crafted code review requests are essential to eliciting quality responses. To address this, we propose a PCR service that leverages the request necessity prediction subtask and the tag recommendation subtask.

When considering a request in the PCR community, we assume the existence of a corpus of PCR requests (denoted by $R$), a set of tag labels (denoted by $L$), and a set of quality labels for all PCR requests (denoted by $Q$).
Formally speaking, given a PCR request $r \in R$, the request contains a title $t \in T$, a text description $d \in D$, and a code segment $c \in C$. Code segments are extracted from descriptions.

\textbf{Request necessity prediction subtask}. 
The request necessity prediction subtask aims to acquire a function $f_{quality}$ that maps $r$ to a quality label $q \in Q$. A review request $r$ corresponds to the truth label $q_{r}$. Thus, the request necessity prediction subtask can be viewed as a classification problem. The target is to identify an optimal function $f_{quality}$ that maps the quality label $q$ as similar to the true quality label $q_{r}$.

\textbf{Tag recommendation subtask}. 
The tag recommendation subtask aims to obtain a function $f_{tagrec}$ that maps $r$ to a subset of tags $l = \{l_{1},l_{2},\dots,l_{m}\} \subset L$, which are most relevant to the request $r$. 
A review request $r$ corresponds to a set of tags $l_r = \{l_{r1}, l_{r2}, \dots, l_{rk}\} \subset l$, where $k$ denotes the number of truth tags for the review request. Thus, the tag recommendation subtask can be viewed as a multi-label classification problem. The target is to identify an optimal function $f_{tagrec}$ that maps the tag subset $l$ as similarly as possible to the truth tag set $l_{r}$.

According to the above definitions, in order to better use the UniPCR to solve these two tasks, we use prompt engineering~\cite{DBLP:conf/icse/BacchelliB13} to reformulate them as generation tasks (details in Section~\ref{frame}). Here, we formally introduce the problem of the PCR service under UniPCR as follows.

\textbf{Our Unified Task.} First, we design a descriptive prompt template for both two subtasks, which is summarized as $T(\cdot) = $``$\textit{x}$ $[MASK]$'' (where $\textit{x}$ denotes a textual string). A request input sequence $r$ is passed through this descriptive prompt and refactored into $r_{prompt} = T(r)$. The request necessity prediction and tag recommendation subtasks is made by filling in the position of $[MASK]$. Our problem is how to improve the proximity of the predicted filler words against the truth label.
% 
% 数据集表示
We denote the total number of training examples as $N$. For the tag recommendation subtask, the truth tags for a request are $l_{r}$. Note that a PCR request can be labeled with up to five tags, so $\left|l_{r}\right| \le 5$.

\setlength{\abovecaptionskip}{0.2cm} 
\section{Framework} \label{frame}

\begin{figure*}[h]
\setlength{\abovecaptionskip}{0.2cm} 
  \centering
  \includegraphics[width=0.9\linewidth]{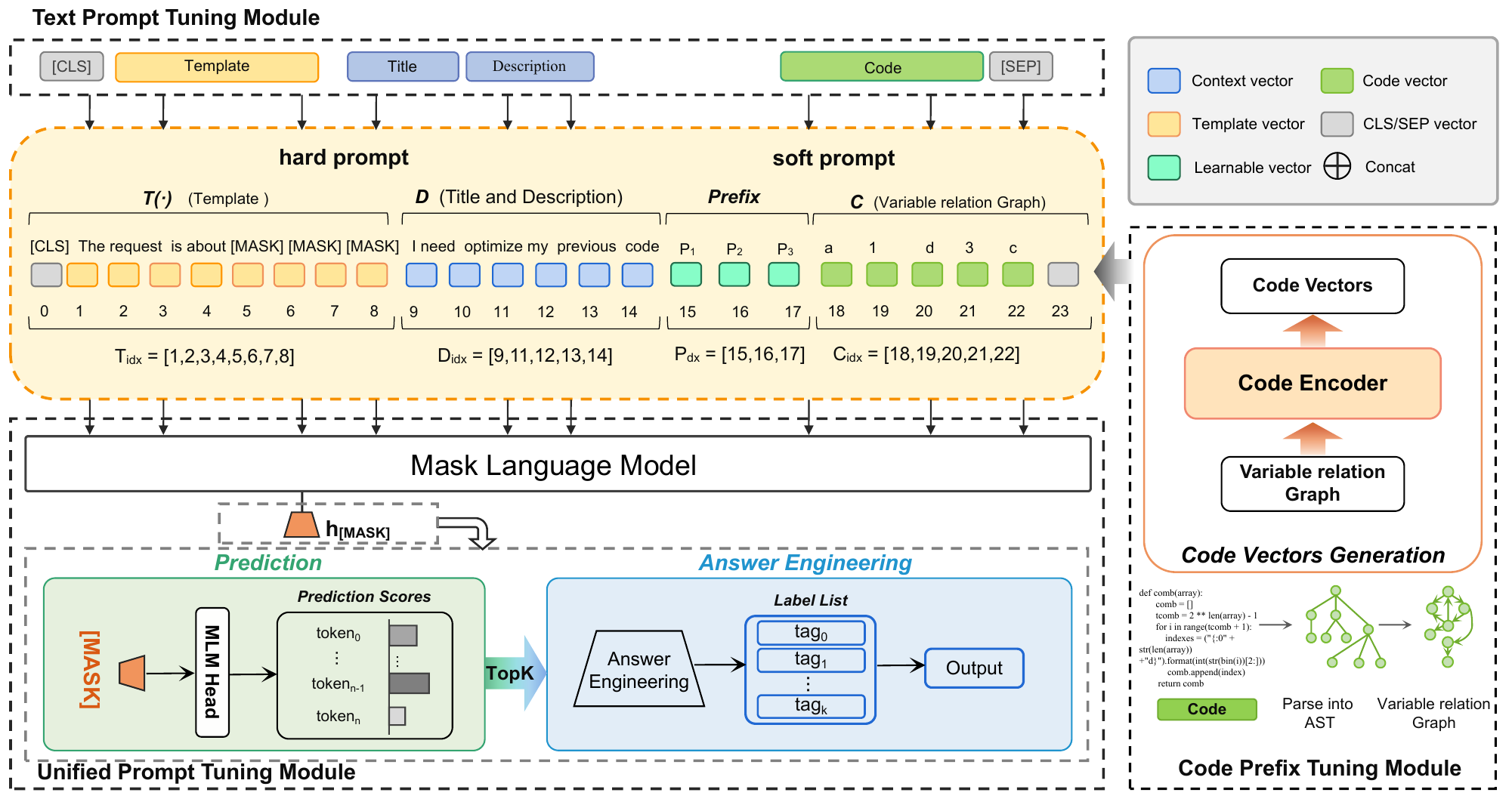}
  \caption{The overview of unified framework for public code review (UniPCR).}
  \label{samples/framework.pdf}
\end{figure*}

Figure~\ref{samples/framework.pdf} plots the architecture of \textbf{UniPCR} for prompt tuning with three modules: the text prompt tuning, the code prefix tuning, and the unified prompt tuning.

In the field of natural language processing (NLP), language prompts have been proven to be effective in leveraging the knowledge learned by pre-trained language models (PLMs)~\cite{DBLP:conf/emnlp/DavisonFR19,DBLP:conf/emnlp/HanGYYLS19}. 
Through a series of research works on knowledge exploration, hard prompt has been used to reformulate downstream tasks with the help of text prompt templates.
By selecting appropriate prompts, the model's behavior can be manipulated so that the PLM can predict the desired output, even without any additional task-specific training~\cite{DBLP:conf/naacl/SchickS21,DBLP:conf/acl/GaoFC20,DBLP:journals/jmlr/RaffelSRLNMZLL20}.

For a request input $r$, we divide the input PCR request into: 1) natural language text $r_{text}$, which is composed of the title and description (text part) in the request, that is, $r_{text} = \{t, d\}$, and
2) code segment $r_{code}$, which indicates the code snippet extracted from the description, that is, $r_{code} = \{c\}$. 
This module uses hard prompt~\cite{DBLP:conf/icse/BacchelliB13} to deal with the input of natural language text by templates and soft prompt for code segment by prefix.

\subsection{Prompt Template Design} 
\subsubsection{Text Prompt}  \label{text prompt section}
First, we apply a descriptive prompt template $T(\cdot)$ to map the input natural language text $r_{text} = \{t, d\}$ to $T(r_{text})$. The prompt template preserves the original markup in $r_{text}$ and includes at least one $[MASK]$ for the PLM to populate with label words as in the two subtasks.

Specifically, we modify the descriptive prompt text string $\textit{x}$ according to the subtask goal to reconstruct the data for different tasks. Expressing $r_{prompt}$ as $T(r_{text})$ = \{\textbf{``The topic of the request is $[MASK]$ $[MASK]$ $[MASK]$. This request necessity is $[MASK]$.''}, $r$\}.

\begin{figure*}[h]
  \centering
  \includegraphics[width=\linewidth]{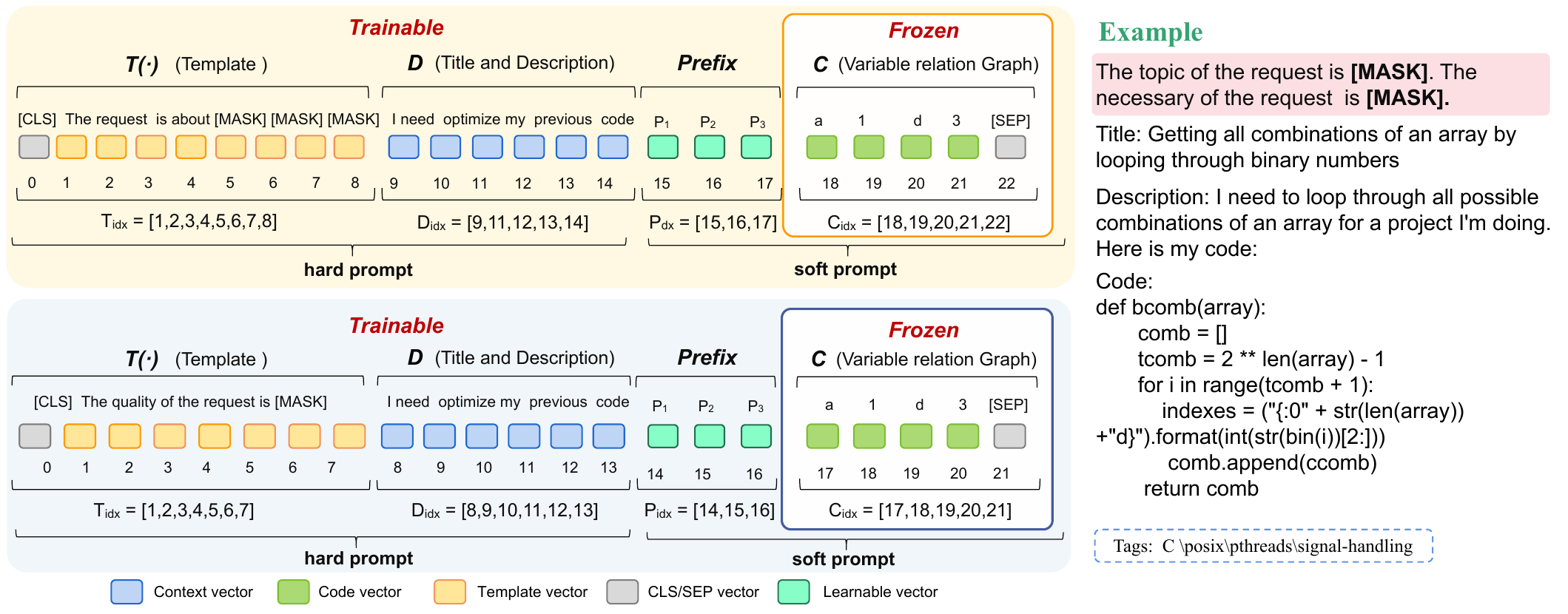} 
  \caption{An Example of Refactoring Request Prediction and Tag Recommendation Subtasks under a UniPCR.}
  \label{prefix}
\end{figure*}

\subsubsection{Code Prompt} \label{code prompt section}
The code segment possesses an inherent structure that provides valuable semantic information for code understanding. To better capture this information, Guo et al.~\cite{DBLP:conf/iclr/GuoRLFT0ZDSFTDC21} proposed using data flow graphs to analyze code and represent its semantic information.
Unlike the Abstract Syntax Tree (AST) used in previous research for code parsing~\cite{DBLP:conf/iclr/HellendoornSSMB20,DBLP:conf/iclr/AllamanisBK18}, this graph-based data flow structure effectively avoids the unnecessary deep hierarchical structure that AST may introduce. 
Following this approach, we parse code segments into data flow graphs to represent their semantics and variable dependent relationships.

To optimize the code representation and better capture the knowledge embedded in the code segment, soft prompt~\cite{DBLP:journals/csur/LiuYFJHN23} is employed. 
Soft prompt is a prompt paradigm that performs directly in the model's embedding space and tunes the prompt parameters to the downstream task. 
This approach has demonstrated superior performance on various non-natural language data, such as image coding and visual question answering~\cite{DBLP:conf/nips/TsimpoukelliMCE21,DBLP:conf/cvpr/LuddeckeE22,DBLP:conf/emnlp/Liu0TC023}. 
It is well-suited for learning and optimizing code segments with such a graph structure, as it can better adapt data forms to a data flow graph.

\subsection{Unified Prompt Tuning in PCR} \label{unified section}

\subsubsection{Prompt Engineering.} 

In Step 1, we first parse a code segment $c$ into a data flow graph to represent the dependency relationship $Graph_{c}$ between code variables (shown on the red part one the right side of Figure~\ref{samples/framework.pdf}). 
%
% In Step 2, the variable relation graph $Graph_{c}$ is sent to a code encoding model pre-trained on code tasks to obtain the hidden vector of the code segment. 

Then the prompt project of prefix tuning is used to refactor the code segment which creates the code input $c_{prefix}$ by obtaining a prefix vector $P_{REFIXcode}$ defined in step 3. To address the issues of low performance and high variance resulting from random initialization of $P_{REFIX}$, it has been suggested that using real words to activate prefix vectors can achieve better results~\cite{DBLP:conf/acl/LiL20}. 
Specifically, we initialize the prefix vector $P_{REFIXcode}$ for code segment prefix tuning using a small piece of code segment as $c'$. 
The small segment vector $P_{REFIXcode}$ is used to optimize its representation based on full-text understanding of code segments, and then a non-updatable data flow graph is adopted based on semantic relations to represent the meaning of individual code segments. This enables a global and local understanding of the meaning of each code segment.

In Step 4, the code segment $c_{prefix}$ replaces the original code segment vectors in the request $r$. Through prompt engineering, the review request is refactored into $r_{prompt}$, and $\tilde{r_{prompt}}$ represents the hidden states of this input. That is, $\tilde{T(\cdot)},\,\tilde{t},\,\tilde{d},\, \tilde{c_{prefix}}$ respectively represent the vectors obtained by $T(\cdot),\,t,\,d,\,c_{prefix}$.

\begin{algorithm}[bt]
	\caption{A Request Prompt Pre-processing}
	\label{curvature}
	\LinesNumbered
	\KwIn{$Request = (R_{1},R_{2} \cdots R_{n})$, each data 
                $R_{i} = r = (t,d,c,l,q)$(title, description, code, tag labels, necessary labels)               }
	\KwOut{$\tilde{r_{prompt}}$}
        {$Graph_{c} \gets  c$ , $Graph_{c}$ represents code data flow graph}\\
        % {${\tilde{c} \gets h_{Graph_{c}}}$, $h_{Graph_{c}}$ represents the hidden vector obtained by the encoding model}\\
        {$P_{REFIXcode} \gets  PLM(c')$, $c'$represents a part of the whole code $c$, $PLM(c')$ represents code embedding by the encoding model\\ }
        {$c_{prefix} = [P_{REFIXcode}, Graph_{c}]$}
        {$\tilde{r_{prompt}} = \{\tilde{T(\cdot)}, \tilde{t}, \tilde{d}, \tilde{c_{prefix}}\}$}	
\end{algorithm}

\subsubsection{Prompt Tuning.} Next, we concatenate the obtained text segment vectors and code segment vectors into the input $\tilde{r_{prompt}}$ of the MLM model. 
In the entire input, $\tilde{Graph_{c}}$ is a non-updatable vector, while the remaining vectors update the representation of the target optimization vector according to the gradient.

% 放到3-3
As shown in Figure~\ref{prefix}, the vector ${\tilde{Graph_{c}}}$ encoded by the hidden layer of code PLM will be frozen and not used for updating, while the prefix is obtained by using the code segment text initialization. Finally, the vector $\tilde{P_{REFIXcode}}$ can be trained.
Code prefix tuning initializes a trainable matrix $P_{\sigma}$ (parameterized by $\sigma$) of dimension $|P_{idx}| \times dim(\tilde{P_{REFIXcode}})$ to store the prefix parameters. 
% The training objective is the same as Equation \ref{eqn-6}. 

As multiple $[MASK]$ tokens may be present in a descriptive prompt template, considering all masked positions for prediction. The \textbf{final learning objective} of our UniPCR framework is to maximize the result of Equation~(\ref{eqn-6}).
\begin{equation}\label{eqn-6} 
\log_{}{p_{\phi}(l|r)} = \prod_{j=1}^{n} p_{\phi}([MASK]_{j} = \varphi_{j}(l_{r})|T(r)).
\end{equation}
where $n$ is the number of masked positions in a descriptive prompt template $T(r)$, and $\varphi_{j}(l_{r})$ is the mapping of the $j$th word in class $l$ and the predicted word to $j$th mask position $[MASK]$.
The set of updatable parameters has changed, and the model parameter $\sigma$ is a fixed parameter. The prefix parameter $\sigma$ is the only trainable parameter. $\tilde{P_{REFIXcode}}$ is a vector that can be used to update $P_{\sigma}$.

\subsubsection{Answering Engineering.} The probability $p([MASK] = y)$ is calculated for each word at position $[MASK]$ by feeding $r_{prompt}$ into an MLM, where $y \in Y$ are tokens in the PLM vocabulary. The probabilities are sorted to obtain a TOP K token list $y_{list}$, which is mapped to the prediction label $P = \varphi(y_{list})$ through the answer project $\varphi(\cdot)$.

To illustrate how we design an answer space, a tag recommendation subtask is used as an example. The answer engineering $\varphi(\cdot)$ will map the predicted word to the final label set $L$ and $Q$. For a given request $r$, a hidden vector $h_{[MASK]}$ can be obtained after passing through the MLM. Subsequently, by passing through the MLM header, a set of probabilities $p_{j}(l|r), j \in (0,\max{|v_{t}|})$ can be retrieved.
By sorting the probabilities, the score for each vocab in the PLM vocabulary $Y$ can be obtained. Since the predicted answer word $v$ may not be exactly the same as the words in the label set, we design a mapping function $\varphi$ to measure the truth of the answer using a matching function and the minimum edit distance. Finally, the mapped answer $\varphi(y)$ is the final top-k output $v_{klist}$.

\section{Experiments} \label{experiments}

\subsection{Dataset and Metrics}

\subsubsection*{\textbf{Dataset.}} 

To evaluate the effectiveness of UniPCR utilizing prompt learning for the PCR (public Code Review) task, we conducted an empirical study using the dataset available which is sourced from StackExchange\footnote{\url{https://archive.org/details/stackexchange}}, the largest public code review website, and it covers the raw data version from 2011 to 2023~\cite{dataset}. In an entire PCR dataset a tag is treated as rare if its occurrence is below a predefined threshold $\theta$. Intuitively, if a tag rarely appears in such a large corpus of data, it is likely to have been a miscreated tag that is not widely recognized by developers. Therefore, these rare tags are removed, as they may reduce the quality of the training and testing data.
In line with previous research~\cite{DBLP:journals/jss/LiXYL20,DBLP:journals/tse/XuHSYXL22,DBLP:conf/iwpc/HeX0HY022}, we set the threshold $\theta$ for rare tags to 50, while all requests and tags that are rare or contain only rare tags are deleted.
In the end, the whole dataset contains a total of 424 tags. After cleaning, we obtained a dataset consisting of 76,161 requests. Following previous work~\cite{DBLP:journals/tse/XuHSYXL22,DBLP:conf/iwpc/HeX0HY022}, the dataset is split into training, validation, and test sets using an 8:1:1 ratio. 

\begin{table}[]
\centering
\caption{Statistics for Public Code Review Dataset (Span 2011-2023)}
\label{dataset}
\begin{tabular}{lccclccc}
\hline
             & Train                     & Test  & Validation   &  & Train                     & Test  & Validation           \\
\hline
Total Data        & 61688    & 7616   &  6855 &     &    &     \\
\hline
Request Necessity       &     &    &   & Tag Recommendation       &     &    &    \\
Necessary          &  26632                            & 3288      &   2959                &Avg. tags/request       &  2.95   & 2.94   &  2.89     \\
Unnecessary & 35056                           & 4328       & 3896           & Avg. words/tag       &  2.89   & 3.0   & 2.89         \\
\hline
\end{tabular}
\end{table}

\subsubsection*{\textbf{Metrics}} For the tag recommendation subtask, previous recommendation tasks~\cite{DBLP:conf/iwpc/HeX0HY022,DBLP:journals/jss/LiXYL20,DBLP:journals/tse/XuHSYXL22,DBLP:conf/wcre/ZhouLYZ17} in the SQA community use $Presision@K$, $Recall@K$, $F1@K$ to evaluate the performance of the approaches. To be consistent, we use the same evaluation metrics in this work.
The request necessity prediction subtask is consistent with the review comment request necessity prediction subtask~\cite{DBLP:conf/sigsoft/Pandya022,DBLP:conf/msr/RahmanRK17} in the PCR field, and $Accuracy$ and the label (necessary for positive and necessary for negative) $F1$ are selected as prediction indicators~\cite{DBLP:conf/sigsoft/Chen0AR0N22}.

\subsection{Implementation Details}

\subsubsection*{\textbf{Implementation}} 
% 即使我们的统一框架可以适用于各种不同的预训练语言模型如：BERT,Roberta,ALBERT,GPT等，但由于硬件设备的限制，BERT被选择作为我们的基准模型。
Even though our UniPCR framework can be applied to various pre-trained language models such as BERT~\cite{DBLP:conf/naacl/DevlinCLT19}, RoBERTa~\cite{DBLP:journals/corr/abs-1907-11692}, ALBERT~\cite{DBLP:conf/iclr/LanCGGSS20}, GPT~\cite{DBLP:conf/nips/BrownMRSKDNSSAA20}, etc. Due to the limitations of hardware devices, BERT was selected as our encoding model (BERT-base-uncased version).
In the UniPCR, an MLM based on BERT is used as the pre-trained model for the experiments below.

Building on the success of PLM in NLP, pre-trained models for programming languages have also been developed. To better capture the inherent structure of code, Guo et al.~\cite{DBLP:conf/iclr/GuoRLFT0ZDSFTDC21} proposed GraphCodeBert, a pre-trained model based on BERT that uses the data flow graph structure of code and natural language. This model shows significant improvements in downstream coding-related tasks. In this work, we leverage GraphCodeBert~\cite{DBLP:conf/iclr/GuoRLFT0ZDSFTDC21} for code model to construct code prefix tuning.

In our UniPCR, the best performance is obtained when the learning rate runs to 1e-5, the initial value of
the trainable curvature to 3.0, then use two hyperbolic graph convolution layers for two datasets. For margin ranking loss, we run the
margin as 1.5 and negative number as 256.  We still use Adam Optimizer \cite{DBLP:journals/corr/KingmaB14} to optimize our model parameters by computing different adaptive learning rates, running the epoch to 6 for the best results, and the model is saved every 10,000 steps. All experiments were performed using version 1.3.1 of Pytorch on a commodity machine equipped with 2 GTX 3090 and a total of 48GB memory.

\subsubsection*{\textbf{Baselines}} 
For overall comparisons, a series of state-of-the-art approaches are chosen as baselines for the tag recommendation subtask and the request necessity prediction subtask, respectively. 

1) For the {\bf{request necessity prediction subtask}}, since existing research pays less attention to request necessity-related tasks, comment-based review work is selected as our baseline experiment~\cite{DBLP:conf/sigsoft/Chen0AR0N22,DBLP:conf/sigsoft/HongTTA22,DBLP:conf/swqd/OchodekSMS22}.

2) For the {\bf{tag recommendation subtask}}, recent methods are selected since they work on recommending tags for the SQA communities~\cite{DBLP:conf/iwpc/HeX0HY022,DBLP:journals/jss/LiXYL20,DBLP:journals/tois/NieLFSWW20,DBLP:journals/tse/XuHSYXL22}.

We reproduce the open source code provided by the authors, re-optimize the parameters for each benchmark on our dataset, and select the best result as the benchmark performance. Since comment services in PCR are often implemented as tools, our replication is based on their published papers. 

\section{Experimental Results}  \label{experimentre}
\subsection{Request Necessity Prediction Subtask}

Table \ref{table2} shows the UniPCR experimental results on the request necessity prediction compared with similarity-based and fine-tuning-based methods.

\begin{table}
\setlength{\abovecaptionskip}{0.2cm} 
\centering
\tabcolsep=0.02\linewidth
 \caption{Experiment Results of Request Necessity Prediction Subtask}
\label{table2}
\begin{tabular}{lcccc} 
\hline
Model  &  Accuracy & F1 Necessary  &  F1 Unnecessary \\ 

   \hline
CommentBERT ~\cite{DBLP:conf/swqd/OchodekSMS22}     & 0.608    & 0.606   &  \underline{0.610}  \\
MetaTransformer~\cite{DBLP:conf/sigsoft/Chen0AR0N22}         & \underline{0.623}      & 0.687  &  0.525  \\
% TagDC         & 0.602      & 0.667   &  0.508   \\
Commentfinder~\cite{DBLP:conf/sigsoft/HongTTA22}        & 0.616      & \underline{0.711}   &  0.426   \\
\hline
\textbf{Our UniPCR}     & \textbf{0.798 }     & \textbf{0.817}   &  \textbf{0.775}   \\

\% Improv.     &  28.0\%   & 14.9\%  &  27.0\%  \\
\hline
\end{tabular}
\end{table}

1) \textbf{Our unified framework is effective on the request necessity prediction subtask.}
PCR requests and comments typically consist of both text and code, three baselines are used to compare the performance of our UniPCR. Those baselines apply traditional finetuning training and similarity paradigm for request necessity prediction downstream subtask. Our unified prompt tuning framework and achieve an improvement of  18.9\% - 28\% over the best performance of baselines. 
In general, UniPCR demonstrates its effectiveness on this subtask by refactoring it into a framework approach such as a generation task.
Through this framework, the model learns label semantics and associates it with the overall information of the request,  resulting in high-quality performance in request necessity prediction.

2) \textbf{UniPCR performs better than traditional architecture.}
Specifically, the baseline~\cite{DBLP:conf/sigsoft/Chen0AR0N22} shows sub-optimal performance, which uses fine-tuning paradigm with a encoding layer for text embedding and a mapping layer for predicting. Our UniPCR framework based on prompt learning has an improvement of 18.9\% over it.
This result highlights that while fine-tuning paradigms can be designed for specific tasks, a more unified framework can achieve competitive results without modifying the internal architecture for specific tasks.
The UniPCR outperforms these different paradigms by 34\%, 18.9\%, and 14.9\% on the ``unnecessary'' F1-score, respectively. Similarly, the UniPCR also shows good performance on the ``necessary'' F1 score 34.8\%, 18.9\% and 14.9\%. Overall, our proposed framework is able to achieve good performance predictions on request necessary by using a unified training of text prompt tuning and code prefix tuning.

\subsection{Tag Recommendation Subtask}

Tables \ref{table1} shows our experimental results on tag recommendation subtask. Comparing the baselines of fine-tuning training methods based on CNN and GraphCodeBERT as feature extractors, our UniPCR framework achieved better performance.

\begin{table}[]
\setlength{\abovecaptionskip}{0.2cm} 
\centering
\tabcolsep=0.02\linewidth
\caption{Experiment Results of Tag Recommendation Task}
\label{table1}
\resizebox{1.0\linewidth}{!}
{
\begin{tabular}{l|ccc|ccc|ccc}
%\toprule
\hline
\multirow{2}{*}{Model}  & \multicolumn{3}{c|}{@3}                                                              & \multicolumn{3}{c|}{@5}                                                              & \multicolumn{3}{c}{@10}                                                             \\ \cline{2-10} 
                              &   \multicolumn{1}{c}{Precision} & \multicolumn{1}{c}{Recall} & \multicolumn{1}{c|}{F1} & \multicolumn{1}{c}{Precision} & \multicolumn{1}{c}{Recall} & \multicolumn{1}{c|}{F1} & \multicolumn{1}{c}{Precision} & \multicolumn{1}{c}{Recall} & \multicolumn{1}{c}{F1} \\\hline%\midrule

TagDC~\cite{DBLP:journals/jss/LiXYL20}                     & 0.429                         & 0.483                      & 0.434                   & 0.311                         & 0.570                      & 0.387                   & 0.189                         & 0.678                      & 0.288                  \\
PORFIT~\cite{DBLP:journals/tois/NieLFSWW20}                    & 0.511                         & 0.571                      & 0.516                   & 0.368                         & 0.668                      & 0.456                   & 0.217                         & 0.774                      & 0.330                  \\
Post2Vec~\cite{DBLP:journals/tse/XuHSYXL22}                & {0.550}                         & {0.624}                      & {0.558}                   & {0.384}                         & {0.707}                      & {0.478}                   & {0.223}                         & {0.799}                      & {0.339}                  \\
PTM4Tag~\cite{DBLP:conf/iwpc/HeX0HY022}                   & 0.488                         & 0.557                      & 0.496                   & 0.339                         & 0.628                      & 0.423                   & 0.199                         & 0.720                      & 0.303                  \\
Dual~\cite{DBLP:journals/inffus/LiWZXTV23}                  & \underline{0.564}                         & \underline{0.635}                     & \underline{0.572}                    & \underline{0.396}                        & \underline{0.724}                     & \underline{0.491}                   & \underline{0.227}                       & \underline{0.806}                      & \underline{0.345}                  \\
\hline
\multicolumn{1}{l|}{\textbf{Our UniPCR}} & \textbf{0.610}                         & \textbf{0.688}                      & \textbf{0.619 }                  & \textbf{0.436}                         & \textbf{0.792}                      & \textbf{0.541}                   & \textbf{0.245}                         & \textbf{0.870}                      & \textbf{0.372}                  \\
\multicolumn{1}{l|}{\%Improv.} &     8.2\%                    &          8.3\%            &       8.3\%            &     10.1\%                     &       9.3\%              &    10.2\%                &                  7.9\%     &  8.0\%                     &        7.8\%         \\

\hline%\bottomrule
\end{tabular}
}
\end{table}

1) \textbf{Experimental results show the effectiveness of the unified framework on the tag recommendation subtask.}
In Table \ref{table1}, our UniPCR framework is compared with the state-of-the-art methods~\cite{DBLP:journals/jss/LiXYL20,DBLP:journals/tse/XuHSYXL22,DBLP:conf/iwpc/HeX0HY022,DBLP:conf/wcre/ZhouLYZ17} applied in the SQA community. Experimental results on the public Code Review dataset demonstrate that our unified prompt tuning method outperforms the best performance of baselines by at least an 11\% - 13.5\% improvement in terms of these metrics.

2) \textbf{Comparison of UniPCR and traditional network architecture on this task.} 
Our UniPCR framework based on prompt learning outperformed the sub-optimal results in @3, @5, and @10, with improvements of 8.2\%, 10.1\%, and 7.8\% in F1, 8.2\%, 10.1\%, and 7.9\% in terms of Precision rate, and 8.3\%, 9.3\%, and 7.9\% in terms of Recall, respectively. Comparing different frameworks, the sub-optimal model Dual add an similarity part with encoding part for tag recommendation. Post2Vec and PTM4Tag adopt a framework that encodes text, code and tags separately.
It can be observed from the experimental results that the UniPCR outperforms the existing two types of frameworks based on the fine-tuning paradigm in terms of task performance. The improvements can attribute to the UniPCR can better learn tag semantics and code representation through soft and hard prompts used in the unified framework.
Overall, our UniPCR framework has better performance on the tag recommendation subtasks than other frameworks.

\subsection{Ablation Study}

To investigate the effectiveness of text prompt tuning and code prefix tuning, an ablation study is conducted to evaluate the contribution of each component. Specifically, we removed the code prefix tuning and the text prompt tuning modules individually.
Since our experiment uses continuous prompt to enhance representation learning of code segments based on the reconstruction task, using continuous prompt to make predictions while abandoning discrete prompt in the experiment would go against the basis of our experiments.

\subsubsection*{\textbf{W/O Code Prefix.}} 
\textbf{The results of code prefix ablation experiments show that this part makes a certain contribution to both subtasks.}
For the code segment prefix tuning ablation study, we modified the input sequence vector by splicing the code segment after the text input and generating the representation of the loss optimization code segment generated at $[MASK]$ positions. This resulted in a new input sequence vector,
\begin{equation}
    \tilde{r'_{prompt}} = \{\tilde{T(\cdot)}, \tilde{t}, \tilde{d}, h_{Graph_{c}}\}, T(\cdot) = \{T_{t}(\cdot) , T_{q}(\cdot)\}
\end{equation}
Then the vector is used as input to the MLM to obtain the final predicted label, using the same inference module as shown in Figure~\ref{samples/framework.pdf}.

We investigated the impact of code prefix tuning on tag recommendation and request necessity prediction tasks. Our results indicate that removing this component leads to a significant degradation in performance. 
Specifically: 
1) In the tag recommendation task, $Precision@k$, $Recall@k$, and $F1@k$ are dropped by 6.9\%-7.4\%, 2\%-2.5\%, and 5.6\%-5.8\%, respectively. 
2) In the request necessity prediction task, $Accuracy@k$, $High\, F1@k$, and $Low\, F1@k$ also dropped by at least 6.6\%, 5.6\%, and 8.1\%, respectively. 

Our results indicate that optimizing code representations, while preserving code segments but not using prefixes, leads to a drop in metrics. This suggests that learning only the representation of code segments in the overall content weakens the understanding of the differences between code segments in different requests. By removing prefixes, the semantic angle of code segments is reduced, making it more difficult to distinguish content representations between similar tags. Consequently, code prefix tuning is essential for capturing the request content and enhancing the ability to differentiate between similar tags.

In the end, the results of the ablation analysis above suggests that code prefixing is beneficial for understanding the representation of code segments.

\subsubsection*{\textbf{W/O Text \& Code Prompt.}}
\textbf{The ablation experiment results show that text prompt tuning in UniPCR has a greater contribution on the two subtasks.}
Based on the findings discussed so far, the text prompt tuning is removed and we evaluate the two tasks using the fine-tuning paradigm.
In order to be consistent with our method, we use BERT as the pre-training model.
Tables~\ref{ab1} and~\ref{ab2} show the results of ablation studies on the tag recommendation and request necessity prediction tasks.

\begin{table}[]
\setlength{\abovecaptionskip}{0.1cm} 
\centering
\caption{Ablation Experiments of Request Necessity Prediction}
\label{ab2}
\tabcolsep=0.01\linewidth
%\begin{tabular*}{\linewidth}{@{}@{\extracolsep{\fill}}c|ccc@{}}
\begin{tabular}{lcccc}
\hline
%\toprule
{Models} &  {Accuracy}    & {F1 Unnecessary}    & F1 Necessary & {$\downarrow$ }  \\ \hline
\textbf{Our UniPCR}              & \multicolumn{1}{c}{\textbf{0.798}} & \multicolumn{1}{c}{\textbf{0.817}} & \textbf{0.775} &{-}\\ 
W/O Code Prefix            & \multicolumn{1}{c}{0.745} & \multicolumn{1}{c}{0.771} & 0.712 &{5.6-8.1\%} \\ 
W/O Code \& Text Prompt             & \multicolumn{1}{c}{0.622} & \multicolumn{1}{c}{0.643} & 0.595 &{21.2-23.2\%} \\  
\hline
%\bottomrule
\end{tabular}
\end{table}

Specifically: 1) The results on the tag recommendation subtask indicate that the fine-tuning is not able to fully utilize all the knowledge learned in pre-training, resulting in a significant drop in performance of at least 16\% on the overall index.
2) The request necessary subtask uses fine-tuning to predict necessary labels by learning sentence-level representations of requests, and the results show that using fine-tuning reduces the overall metric by at least 14.9\%. 

This result highlights the importance of prompt tuning in reducing the gap between downstream tasks and the pre-trained model. By reconstructing subtasks to use the vector of the mask position to learn the representation of tags, our framework strengthens the understanding of tag semantics compared to fine-tuning. Additionally, the framework considers the association between tag semantics and request content, further improving the representation learning.

\begin{table}[]
\setlength{\abovecaptionskip}{0.2cm} 
\centering
\caption{Ablation Experiments of Tag Recommendation}
\label{ab1}
\tabcolsep=0.03\linewidth
%\begin{tabular*}{\linewidth}{@{}@{\extracolsep{\fill}}c|ccc@{}}
\begin{tabular}{l|cccc}
%\toprule
\hline
\multirow{2}{*}{Models} & \multicolumn{4}{c}{Precision@k}                                   \\ \cline{2-5} 
                        & \multicolumn{1}{c}{@3}    & \multicolumn{1}{c}{@5}    & @10 & {$\downarrow$ }  \\ \hline
\textbf{Our UniPCR}              & \multicolumn{1}{c}{\textbf{0.610}} & \multicolumn{1}{c}{\textbf{0.436}} & \textbf{0.245} &{-}\\ 
w/o Code Prefix           & \multicolumn{1}{c}{0.565} & \multicolumn{1}{c}{0.404} & 0.228 & {6.9-7.3\%} \\ 
w/o Code \& Text Prompt             & \multicolumn{1}{c}{0.498} & \multicolumn{1}{c}{0.354} & 0.208 & {15.1-18.8\%}\\ 
\hline
\multirow{2}{*}{Models} & \multicolumn{4}{c}{Recall@k}                                      \\ \cline{2-5} 
                        & \multicolumn{1}{c}{@3}    & \multicolumn{1}{c}{@5}    & @10 & {$\downarrow$ }  \\ \hline

\textbf{Our UniPCR}              & \multicolumn{1}{c}{\textbf{0.688}} & \multicolumn{1}{c}{\textbf{0.792}} & \textbf{0.870} &{-}\\ 
w/o Code Prefix            & \multicolumn{1}{c}{0.671} & \multicolumn{1}{c}{0.772} & 0.853 & {2.0-2.5\%}\\ 
w/o Code \& Text Prompt               & \multicolumn{1}{c}{0.586} & \multicolumn{1}{c}{0.673} & 0.773 & {11.1-15.0\%} \\ 
\hline
\multirow{2}{*}{Models} & \multicolumn{4}{c}{F1@k}                                          \\ \cline{2-5} 
                        & \multicolumn{1}{c}{@3}    & \multicolumn{1}{c}{@5}    & @10  & {$\downarrow$ } \\ \hline

\textbf{Our UniPCR}             & \multicolumn{1}{c}{\textbf{0.619}} & \multicolumn{1}{c}{\textbf{0.541}} & \textbf{0.372} &{-} \\ 
w/o Code Prefix            & \multicolumn{1}{c}{0.583} & \multicolumn{1}{c}{0.510} & 0.351 & {5.6-5.8\%}\\ 
w/o Code \& Text Prompt              & \multicolumn{1}{c}{0.513} & \multicolumn{1}{c}{0.446} & 0.319 & {14.2-17.6\%}\\ 
%\bottomrule
\hline
\end{tabular}
\end{table}

This finding highlights the importance of reconstruction tasks through text prompt tuning to achieve better performance and narrow the gap between downstream tasks and pre-trained models. The UniPCR has consistently delivered effective performance across both subtasks (tag recommendation and request necessary), indicating its potential to be applied to other PCR subtasks. These results demonstrate the effectiveness of our approach and its potential for wider applications in the field of public code review.

\section{Related Work} \label{related}
\textbf{Public Code Review} (PCR) is a widely used software engineering practice in which developers manually examine code written by their publics~\cite{DBLP:conf/icse/BacchelliB13,DBLP:journals/software/MacLeodGSBC18}.
Many mature and successful Open Source Software (OSS) projects have recently integrated public code review as a crucial quality assurance mechanism \cite{DBLP:conf/esem/BosuC13}.

1) Request Necessity Prediction Subtask.
Hong et al.~\cite{DBLP:conf/sigsoft/HongTTA22} propose that even though deep learning models such as T5~\cite{DBLP:journals/jmlr/RaffelSRLNMZLL20} achieve impressive performance in the PCR domain, they require expensive computing time and resources, which most software organizations may have limited access to~\cite{DBLP:conf/sigsoft/FuM17}.
To this end, a simple and fast similarity-based retrieval model is proposed to recommend reviews that may be written by reviewers to reduce manual costs. 
Chen et al.~\cite{DBLP:conf/sigsoft/Chen0AR0N22} argue that promoting better CRs requires understanding requests and their quality.

2) Tag Recommendation Subtask.
Tag recommendation technology has become a research hotspot for better serving the learning and communication of software information stations, leading to numerous research works focusing on software information stations ~\cite{DBLP:journals/jss/LiXYL20,DBLP:journals/tse/XuHSYXL22,DBLP:conf/iwpc/HeX0HY022,DBLP:conf/wcre/ZhouLYZ17}. However, some existing research would remove the code segment, assuming that it was noise that affected the final model.
For instance, Zhou et al.~\cite{DBLP:conf/wcre/ZhouLYZ17} proposed TagMulRec, a tool that recommends post tags for developers based on index calculation similarity. 
Li et al.~\cite{DBLP:journals/jss/LiXYL20} proposed TagDC, a composite tag recommendation method with deep learning and collaborative filtering.
In recent years, Xu et al. proposed Post2Vec~\cite{DBLP:journals/tse/XuHSYXL22}, a tag recommendation method based on deep learning that incorporated an understanding of code segments into the model for the first time. He et al.~\cite{DBLP:conf/iwpc/HeX0HY022} incorporated PLMs into the software community domain and used a feature extractor for each piece of content (title, description, and code).

\textbf{Prompt tuning} can be roughly categorized into two groups: hard prompt is constructed from natural language, and soft prompt is from continuous vectors. Soft prompt, such as the method proposed by Liu et al.~\cite{DBLP:journals/corr/abs-2110-07602} called P-Tuning v2, use trainable continuous vectors at each layer to increase the deep capability of soft prompt. Similarly, Li et al.~\cite{DBLP:conf/acl/LiL20} proposed Prefix-tuning, a lightweight alternative that optimizes a small vector of successive task-specific parameters, achieving similar performance by learning only 0.1\% of the parameters compared to fine-tuning. 
Hard prompts are more interpretable because they use natural language as a prompt template. Han et al.~\cite{DBLP:journals/aiopen/HanZDLS22} generated a set of rules called PTR, which decomposes a difficult task into simpler subtasks. Sun et al.~\cite{DBLP:conf/coling/SunZHQ22} proposed NSP-BERT, a method that refactors multiple tasks into NSP form.

Existing research focuses on tasks from the reviewer's perspective in team code review, and needs to design an architecture for each task separately. In the field of public code review, request quality depends more on developers. To complete request quality assurance more simply and quickly, we propose a framework to unify the training objectives of the request necessity prediction subtask and tag recommendation subtask.

\section{Conclusions and Future Work} \label{conclusion}
This work proposes the UniPCR framework to jointly consider both developers and reviewers in the PCR service. 
We consider the request necessity prediction and tag recommendation subtasks in public code review process, and demonstrate how these two tasks serve the process.
Experimental results show that our unified prompt tuning significantly outperforms state-of-the-art methods in both subtasks. 
Additionally, we conducted a series of ablation studies to investigate the effectiveness of text and code prompts. Finally, the UniPCR framework is promising to be transferable to other public code review tasks such as comment request necessity prediction task. 
In the next step, since the UniPCR framework can also be adapted to Large Language Models (LLM), its performance on LLMs will be a promising direction for further exploration.

\section*{Acknowledgments}
The research is supported by National Natural Science Foundation of China: No. 62276196, 52031009 and 62172311, and  the Guangxi Science and Technology Major Program (Guangxi New Energy Vehicle Laboratory Special Project: AA23062066).

\bibliographystyle{splncs04}
\bibliography{mybib}

\end{document}